\documentclass{aa} 
\usepackage{graphicx} 
\begin{document}

   \title{Gravitational instability and clustering in a disk of planetesimals}

   \author{
P. Tanga 
          \inst{1}\fnmsep\thanks{PT acknowledges the support of the
        Poincar\'e fellowship of the Observatoire de la C\^ote
        d'Azur and of the Programme National de Plan\'etologie.} 
          \and 
      S.J. Weidenschilling\inst{2}
  \and
        P. Michel\inst{1}
  \and 
      D.C. Richardson\inst{3} 
          } 
 
   \offprints{P. Tanga, \email{tanga@obs-nice.fr} }
 
   \institute{
    Laboratoire Cassiop\'ee CNRS/UMR6202, Observatoire de la C\^ote d'Azur - BP
    4229 - 06304 Nice Cedex 04, France 
     \and 
     Planetary Science Institute, 1700 East Fort Lowell Road, Suite
    106 - Tucson, AZ 85719-2395 USA
           \and 
             Dept. of Astronomy, Univ. of Maryland, College Park MD  
         20742-2421, USA
             } 
   \date{Received ; accepted} 
 
   \abstract{For a long time, gravitational instability in the
     disk of planetesimals has been suspected to be the main
     engine responsible for the beginning of dust growth, its
     advantage being that it provides for rapid growth.  Its real
     importance in planetary formation is still debated, mainly
     because the potential presence of turbulence can prevent the
     settling of particles into a gravitationally unstable layer.
     However, several mechanisms could yield strongly inhomogeneous
     distributions of solids in the disk: radial drift, trapping in
     vortices, perturbations by other massive bodies, etc.  In this
     paper we present a numerical study of a gravitationally unstable
     layer.  This allows us to go beyond the classical analytical
     study of linear perturbations, exploring a highly non-linear
     regime.  A hierarchical growth of structure in the presence of
     dissipation (gas drag) can yield large, virialized clusters of
     planetesimals, the first time such clusters have been observed in
     the context of planetesimal disks.   }
   \titlerunning{Gravitational instability in planetesimal disks}
   \maketitle
%
 
\section{Introduction}

The formation of planetesimals from dust in the solar nebula, and
their growth to form bodies of planetary size, is a complex process.
In the earliest stages of this process, dust-gas interactions,
collective effects, and timescales are topics of intense research.
Controversy persists as to whether the first solid bodies grew solely 
by physical collisions and sticking, by gravitational instability in
a layer of dust (or larger particles), or by some combination of 
these processes.  

The first detailed qualitative description of planetesimal formation 
was due to K. E. Edgeworth, who is better known for having postulated 
the existence of a huge reservoir of cometary bodies beyond Neptune's
orbit (the so-called Edgeworth-Kuiper Belt).  He inferred that
the settling of particles toward the central plane of a disk-shaped
nebula would produce a layer with density much greater than that of 
the gas, with sufficient density to become unstable due to 
self-gravity (Edgeworth 1949).  In order for this instability to 
occur, the mutual attraction of the particles had to overcome both 
the tidal pull of the Sun and the "internal heat" associated
with their velocity dispersion.  The resultant clusters of particles
could then collapse into solid bodies, which would collide and coalesce 
into larger and larger bodies, eventually producing the planets.

The first mathematical analyses of this process were produced by V. S. 
Safronov (1969), and independently by Goldreich and Ward (1973).  They
performed linear stability analyses of a particle layer with keplerian
rotation. They showed that the particle velocity dispersion stabilized
density perturbations of small spatial scale, while rotation acted to
stabilize large ones.  That is to say, there was a preferred scale for 
gravitational instabilities, which would tend to produce condensations
with a characteristic size.  These analyses assumed that there was no
obstacle to settling of particles, so that the particle layer could
attain the critical density.  However, the presence of any kind of
turbulence in the gas, as is generally accepted as an unavoidable source
of viscosity in accretion disks, could inhibit sedimentation of small 
particles.  Even in a purely laminar nebula, settling of particles into
a dense layer could be inhibited.  The nebular gas necessarily has a
slightly sub-keplerian velocity due to the existence of a radial
pressure gradient.  If a particle layer in the midplane attains a density
greater than that of the gas, it lacks pressure support and tends toward
more nearly keplerian rotation.  The resulting shear generates turbulence
that may prevent further settling, so that the critical density for
gravitational instability is not reached (Weidenschilling 1980).  However,
the effectiveness of this process is currently under debate (Youdin and
Shu 2002, Weidenschilling 2003).  If the dust/gas ratio can be enhanced 
several times over solar abundance, by global transport (Youdin and Shu 
2002) or local shear instabilities (Goodman and Pindor 2000), it may be
possible to attain the critical density.  It may be achieved more simply
if some degree of particle coagulation occurs, as decimeter-sized or
larger bodies will decouple form shear-induced turbulence.

Attainment of the critical density allows gravitational instability, 
but does not guarantee it.  A realistic system of particles would contain
bodies with a range of sizes; as the gas drag produces size-dependent
radial velocities, there will be some velocity dispersion.  This can 
inhibit instability, until the the mean size becomes larger than a
threshold size, estimated by Weidenschilling (1995) to be of the order of
10 meters.  However, the outcome of gravitational instability in that case
may be significantly different from that experienced in a layer of small
particles.  Goldreich and Ward (1973) assumed cm-sized particles,
which would be subject to damping by collisions and gas drag, and set
the velocity dispersion to zero, allowing instabilities on all length
scales below the limit set by rotation.  In contrast, 10-m bodies are
only weakly damped, and may retain a significant velocity dispersion that
favors instabilities at a single scale length (Weidenschilling 1995).
Numerical simulations of particle coagulation in the outer solar system,
where cometary nuclei presumably accreted, showed that gravitational
instability could have occurred in such an environment (Weidenschilling
1997; henceforth W97).  At 30 AU from the Sun, the size of the unstable
regions would be $\approx 0.016$~AU for a low-mass solar nebula, containing 
$\approx 10^{22}$~g of condensed matter.  Weidenschilling speculated that clusters 
of decameter-sized bodies on this scale would not collapse to solid
bodies with diameters $\approx 100$~km; rather, he suggested that due to the lack 
of damping such condensations would be transient.  However, the detailed 
evolution of such clusters, their duration, and their possible influence 
on the growth of planetesimals have not been explored by numerical
simulations.  In general, two elements play against gravitational 
collapse: the overall velocity dispersion, if not damped by collisions or gas
drag, and the angular momentum of a gravitationally bound cluster due
to the keplerian rotation of the particle layer.  If both of these 
quantities are too large, the first can prevent instability on small
scales, while the second can prevent collapse beyond a certain upper-limit
scale.

As detailed in Sect.~2, all these factors are taken into account in
the linear perturbation analysis of the stability of a nearly
homogeneous disk.  However, the subsequent evolution can be highly
non-linear and its study requires the use of numerical simulations.

Given the long-lasting interest in the general problem of
gravitational instability and its importance for our understanding of
the planetesimal growth process, both in our solar system and
elsewhere, we present in this work the numerical study of the
evolution of a gravitationally unstable layer of planetesimals in
orbit around the Sun.  Simulating the process allows us to explore the
collective evolution, going beyond the linear approximation usually
employed to describe the onset of gravitational instability, and to
study the dynamics of planetesimal clusters.  A fully consistent
simulation of the W97 scenario is, however, well beyond the current
computational limits.  In fact it is simply not possible to simulate a
sufficiently large disk patch at 30~AU with a number of particles,
distributed in a range of sizes, large enough to reproduce the
evolution and accretion of a planetesimal swarm starting from 1~cm or
1~m bodies.

We use the $N$-body hierarchical tree code {\sl pkdgrav},
(\cite{Derek}) which is able to treat physical collision between
particles in a consistent way, simulating their coalescence into
larger bodies.  Given the large numbers of interactions and particles
involved, all the runs presented here required the use of
high-performance parallel computers\footnote{In particular, we wish to
  acknowledge the use of SIVAM-II at the Observatoire de la C\^ote
  d'Azur, of the PC cluster installed in the same institute by
  ALINEOS, and of the IDRIS computing facilities.}.  The unstable
regime was explored both in a gas-free environment and in the presence
of a Keplerian non-turbulent gas disk.  We note that the results
obtained here can also be helpful for understanding the process of
gravitational collapse in other scenarios where gravitational
instability could play a role, such as those invoking particle
concentration in the core of large-scale anticyclonic vortices
(\cite{barge95}, \cite{tanga96}).

In the following section we will address the numerical method
employed, together with boundary conditions.  Later (Sect.\ 3) we
illustrate the evolution of a layer of single-sized large
planetesimals, considering either energy-conserving (perfectly
elastic) mutual collisions or inelastic bouncing.  In Sect.\ 4 we show
the effect of enhancing dissipation by including the gas drag exerted
by a laminar, Keplerian flow.

\begin{figure*}[ht]
   \centering
   \includegraphics[width=13cm]{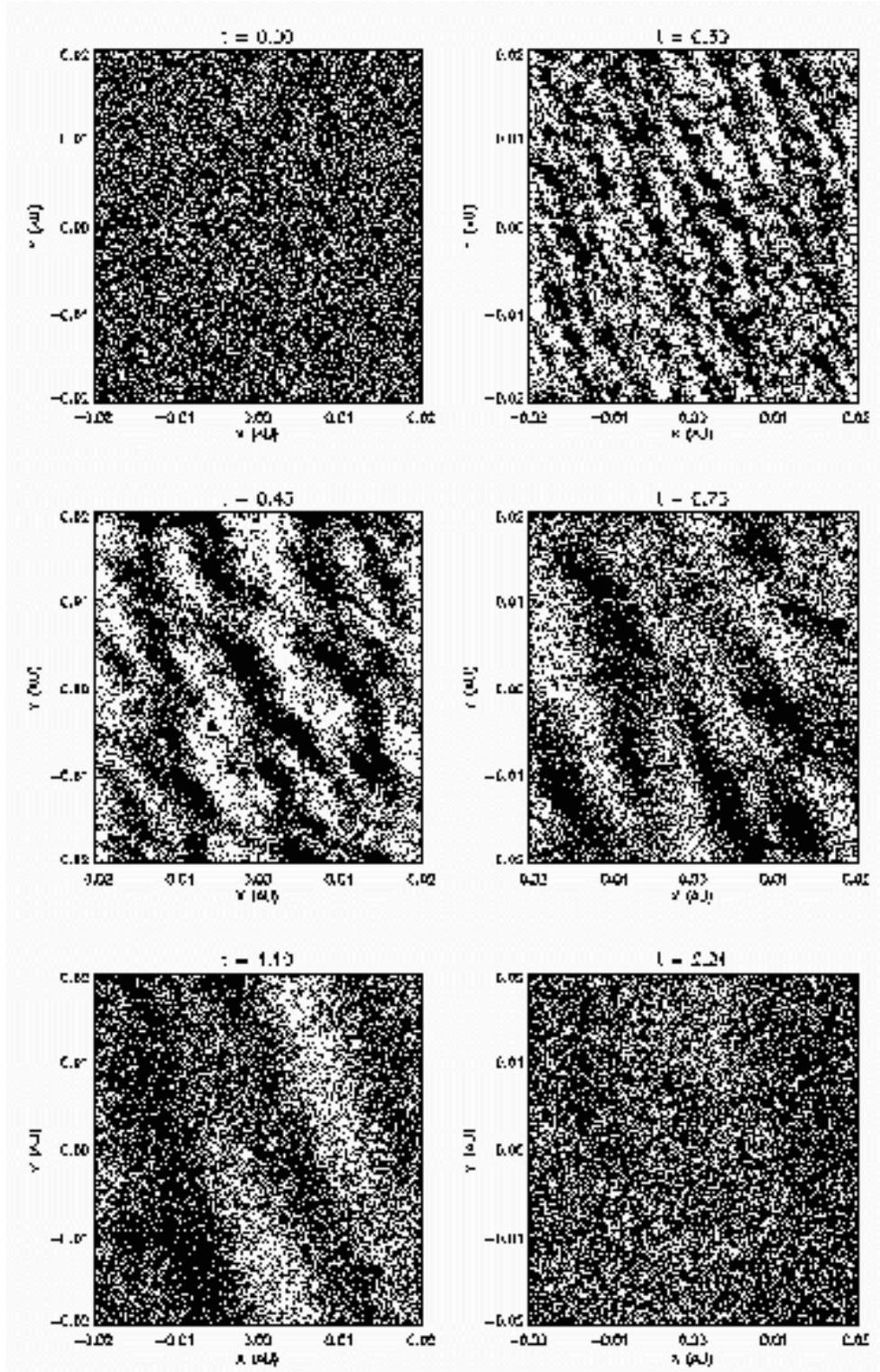}
   \caption{
     Particles positions as projected on the $(x,y)$ plane in the CM
     simulation.  The first panel represents initial conditions.  Time
     is expressed in orbital periods (at 30 AU, about 164 years).
\label{F:nogas_cm1}}
\end{figure*}

\section{Numerical framework}

The {\sl pkdgrav} parallel $N$-body code allows in principle for
treatment of a large number ($N$) of particles, and numerical
experiments have been successfully performed with $N$ ranging from
$10^5$ up to $10^6$ (see for example \cite{Derek}, \cite{tanga02}).
The main advantage provided by this code for our purpose is the
possibility to assign a physical size to particles (considered to be
spherical) and to efficiently detect mutual collisions.  The
collisional outcome can be treated in different ways.  In the
following we will show results obtained either by imposing perfectly
elastic bouncing between the colliders, or by making the colliders
merge into a single body of mass equal to the sum of their masses,
with the same internal density.  Considering the colliding bodies as
an isolated system, the first case implies energy conservation in the
reference frame referred to the center of mass.  In the second case,
dissipation occurs.  Accordingly, we will refer to the two families of
simulations with the acronyms ``EC'' (for ``energy conserving''
collisions) and ``CM'' (for ``collisional merging'').

All the simulations presented in the following have been performed in
the approximation introduced by Wisdom and Tremaine (1988), that is by
assuming that a dense ring of planetesimals at distance $a$ from the
central body can be divided into local boxes (or ``patches'') of size
$L$, with $L \ll a$.  In a coordinate frame centered on the box, with
the $x$-axis pointing away from the Sun, the $y$-axis being tangential
to the direction of motion, and the $z$-axis perpendicular to the
orbital plane, the equation of motion for one particle can be
linearized:
\begin{equation}
  \begin{array}{l}
    \ddot{x} = F_x + 3\Omega^2 x + 2\Omega \dot{y},\\
    \ddot{y} = F_y - 2 \Omega \dot{x},\\
    \ddot{z} = F_z - \Omega^2 z,
  \end{array}
  \label{eqmotion}
\end{equation}
in which $\Omega$ is the angular frequency at distance $a$ from the
Sun, and $F$ represents the gravitational force (per unit
mass) due to the other particles.

The box in which the simulation is performed is considered to be
periodic in the $(x,y)$ plane.  In order to reduce boundary effects,
the gravitational contribution coming from three orders of ghost
patches, each containing a copy of the particles in the main domain,
is considered.  The ghost patches are nested around the central one;
those having a distance from the Sun different from $a$ have a
velocity tangential to the orbit consistent with the Keplerian shear.

It must be noted that the collective particle behavior can be
associated to structures (such as clusters, waves, etc.) whose typical
scale length must be much smaller than $L$ to avoid boundary effects.
However, given the limited resolution available in terms of $N$, it
may happen that adequate sampling of all interesting scales present in
the disk is not possible.  In other words, while the largest scales
are always limited by the size of $L$, the smallest possible scales that can
be represented in the simulation have a size corresponding to several
times the average particle diameter.

The spatial scales that we are interested in can be estimated from the
linear theory of gravitational instability in a rotating disk of
planetesimals (in the following, we use the notation of Ward, 1976).
The dispersion relation that is obtained, in the case of a thin disk
locally rotating at frequency $\Omega$, for a slightly overdense
domain of size $\lambda$, can be written as:

\begin{equation}
  F(\lambda) = 2\pi^2c^2 - 4\pi^2G\Sigma\lambda+\lambda^2\Omega^2 < 0 ,
  \label{dispersion_base}
\end{equation}
in which $c$ represents the velocity dispersion and $\Sigma$ the local
surface density of solids.  An unstable solution exists when
$F(\lambda)<0$, i.e., for $c < c_{crit}=\pi G \Sigma/\Omega$.  The
widest range of unstable wavelengths is obtained when $c=0$; in that
case, for $\lambda<2\lambda_{crit}$ instability is possible.
$\lambda_{crit}$ is the value of the most unstable wavelength (minimum
of the dispersion relation) and is given by:
\begin{equation}
  \lambda_{crit}=\frac{2\pi^2G\Sigma}{\Omega^2}.
  \label{wave_growth}
\end{equation}

As will be shown, the overdense regions grow by hierarchical merging
into larger and larger structures.  However, before their sizes become
comparable to $L$, the increase in velocity dispersion stabilizes the
dust layer, so that these structures dissipate.

\section{Unstable structure growth in a ``cold'' disk}

We simulate here a patch of the disk at an average distance from the
Sun of $a_0 = 30\ AU$.  Since the observed structures develop on
scales $\ll \lambda_{crit}$, a box size comparable to $\lambda_{crit}$
does not appear to be a serious limitation.  In the following, we
choose a patch size $L=0.04$ AU, i.e., $\sim$70\% larger than the
critical wavelength, $\lambda_{crit} \sim 0.024$ AU.  We checked one
case with $L=7\lambda_{crit}$, but no qualitative difference was noted
in the evolution of structures.  We are thus confident that the
results presented here are not seriously affected by boundary effects.

The imposed initial conditions are derived from the comet formation
model in W97 after $t \sim 10^5\ yr$ of model evolution.  At that
stage, a theoretically unstable layer is formed.  This choice is just
a reasonable starting point for investigating the dynamics of
planetesimals in the comet-forming region even though, as we will see,
the limitations in computing power prevent a direct comparison to the
W97 scenario.

The conditions for the onset of the instability require a density
above a certain critical value, so that self-gravity can overcome the
solar tides that otherwise have the tendency to shear away all
structures.  This critical density is usually written as:
\begin{equation}
  \rho_{crit} = \frac{3 M}{2\pi a^3} = \frac{3\Omega^2}{2\pi G} .
\end{equation}

The initial vertical dispersion $h$ chosen in our simulations
corresponds to a volume density $\rho\sim 16\times\rho_{crit}$,
consistent with W97.  The initial positions of the particles are
generated by assigning random positions inside a $L\times L\times h$
box, avoiding particle overlaps.  The components of the velocity
dispersion on the disk plane, that we indicate by $c_x$ and $c_y$, are
initially set equal to zero.  As a consequence, the particles start on
ideal, circular Keplerian orbits.  Concerning $c_z$, its value is
imposed to be consistent with the vertical dispersion.
 
The surface density is $\Sigma = 0.41$ g cm$^{-2}$.  W97 states that
``about two-thirds'' of the total solids take part in the instability.
However, we did not reduce the average surface density accordingly,
preferring to let the dynamics of the layer determine the evolution of
the entire mass.

A good compromise between the simulation speed, the number of
particles and the time step was found for $N=4\times10^4$, with
$1.6\times10^5$ steps per revolution.  This allows for computation of
the dynamical evolution over a few orbits in about 100 hours of CPU on
a cluster of four SP4 processors.

In order to reproduce the surface density given above, with the
material bulk density of particles set to $1\ g\ cm^{-3}$, the
particle radius must be taken equal to $R$=10~km.  This size is about
100 times the particle size that is built at $t \sim 10^5\ yr$ in the
model of W97.  The consequence is a stronger gravitational stirring:
we can thus expect that, in the absence of any dissipation, the
vertical dispersion of the planetesimal disk will rapidly increase
beyond the initially imposed value.

Because of this large size, it is clear that we will not be able to
obtain a realistic simulation of the comet-forming region.  A direct
comparison with the scenario of W97 is thus not straightforward.
Nevertheless, we can infer some interesting properties of the
collapsing layer, and shed some light on previously ignored behaviors.

\begin{figure}[h]
  \centering
  \includegraphics[width=8.5cm]{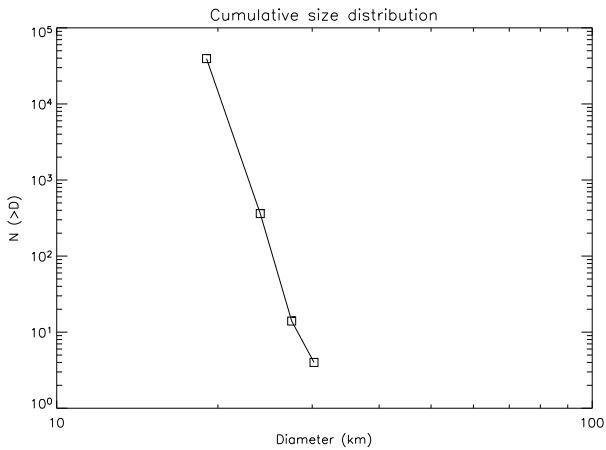}
  \caption{ 
    Cumulative size distribution for the CM simulation at the end of
    the run ($t$=2.24 orbits).
    \label{F:sizedist_cm1}
  }
\end{figure}

In this case, the EC and the CM simulations have several features in
common.  According to (\ref{wave_growth}), the small wavelengths
emerge first (Fig.~\ref{F:nogas_cm1}).  The growth of structures at
larger scales is associated with the gradual coalescence of larger and
larger structures, similar to what happens during the process of
hierarchical growth already observed in numerical simulations of
cosmological structures.  The main difference with cosmology, here, is
the deformation of clusters due to shear, translating into the
elongated structures that in the context of planetary rings are known
as ``wakes'' (see e.g., \cite{derek94}).  Later, the structures
dissipate at all scales.  This sequence of events is evident in
Figure~\ref{F:nogas_cm1}, which shows snapshots from the CM case.  The
instability, here, is not sufficient to create conditions for an
efficient local collapse, and the size distribution of the CM
simulation (Fig.~\ref{F:sizedist_cm1}) shows that only a small
fraction of the population has moderately grown in size.

\begin{figure}[h]
  \centering
  \includegraphics[width=8.5cm]{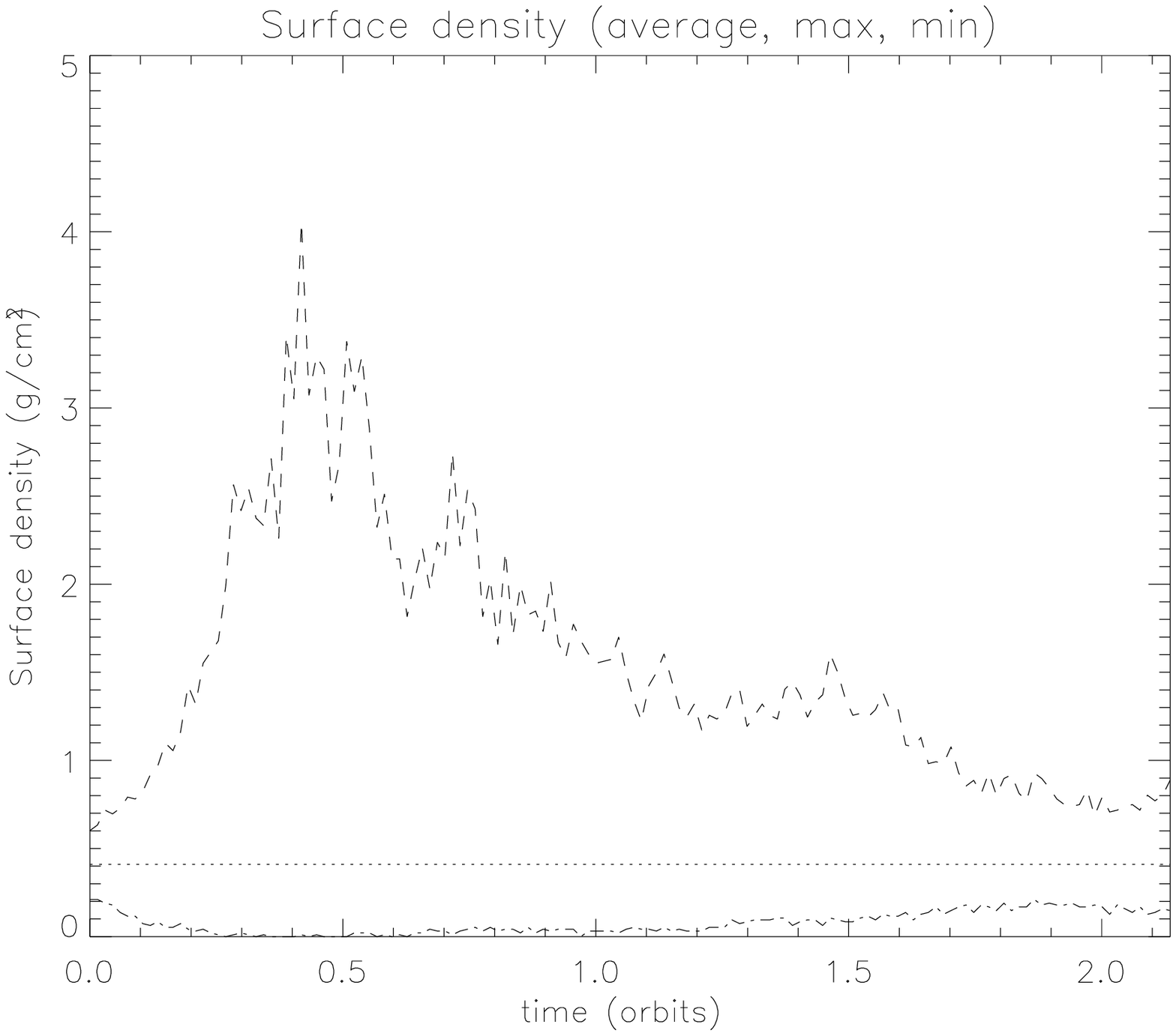}
  \includegraphics[width=8.5cm]{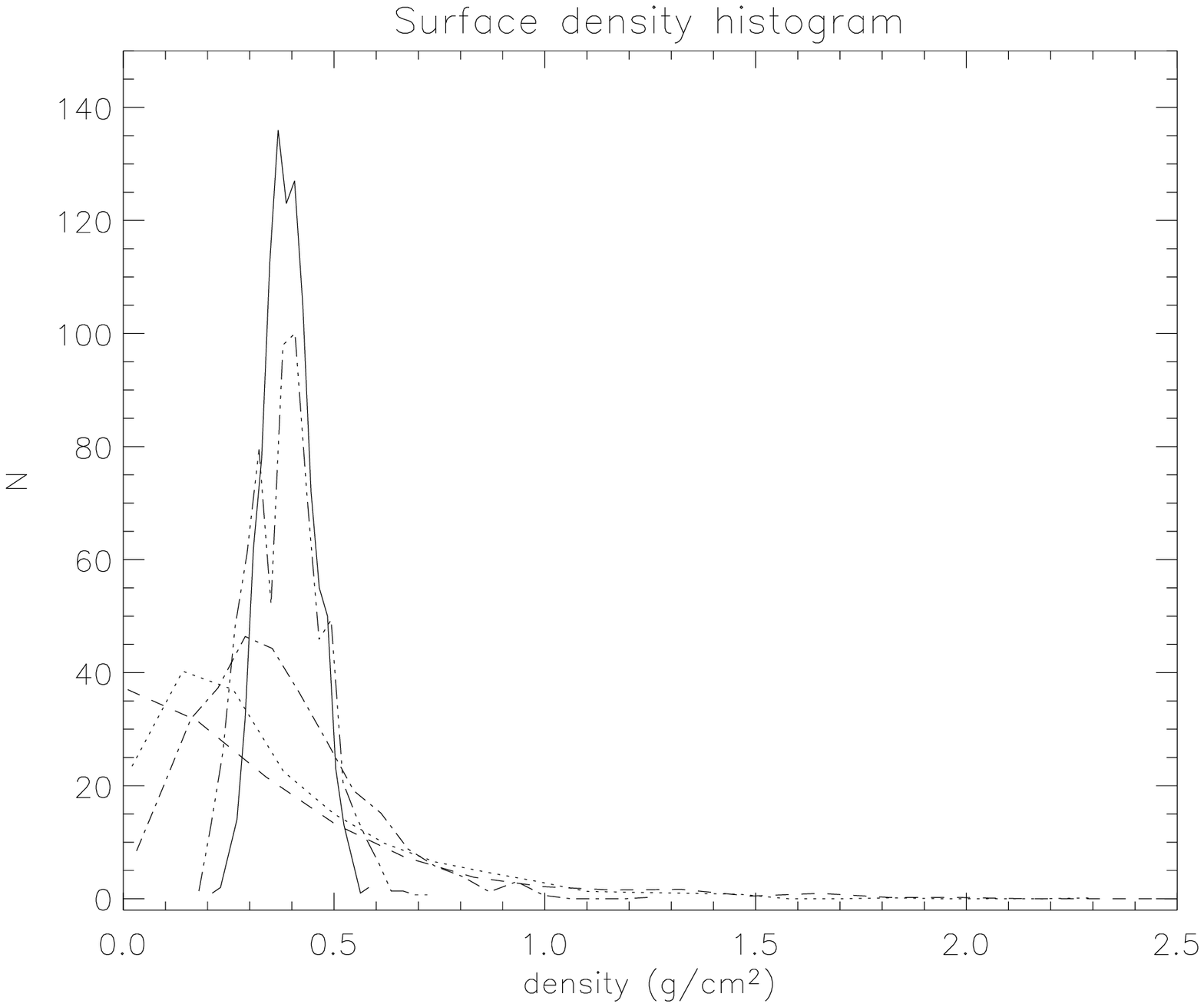}
  \caption{
    The two panels show the time evolution of the local surface
    density, computed on a regular grid of 32$\times$32 boxes, in the
    simulation of Fig.~\ref{F:nogas_cm1}.  The upper panel shows the
    evolution in time of the maximum and the minimum value; the
    horizontal dotted line corresponds to the average.  The lower
    panel shows the histogram of the density values at different
    times.  The continuous line refers to the initial conditions ($t$
    = 0).  The dashed line (with maximum at 0) refers to $t$ = 0.30
    orbit, the dotted to $t$ = 0.45 orbit, the dash-dotted to $t$ =
    0.74 orbit.  The remaining line, at the end of the evolution ($t$
    = 2.25 orbits) approaches again the original Gaussian
    distribution, centered on the average value.
    \label{F:densevol_cm1}
  }
\end{figure}

In order to get some quantitative estimate of the local density
enhancement, we computed the surface density $\Sigma(n,m)$ on a $N
\times M$ grid ($N$ being the dimension along $x$).
Figure~\ref{F:densevol_cm1} (upper panel) shows the time evolution of
the maximum density found in the domain, using $N=M=32$.  A peak
corresponding to a ten-fold enhancement can be identified at $t\sim
0.42$~orbit.  As expected from visual inspection, at later times the
distribution slowly recovers its initial homogeneity.  This can be
seen also in a histogram of the density at different times
(Fig.~\ref{F:densevol_cm1}, lower panel), which shows that the rapid
initial clustering creates strongly overdense regions and a large
population of nearly empty areas (dashed line).  Later, the peak of
the distribution gradually returns toward the average value and the
tail at large densities is lost.

An estimate of the wavelengths of the structures at given times can be
obtained by computing the power spectrum of the density distribution.
In principle, this could be derived from the 2-D Fourier transform
(FT) of the surface density field $\Sigma(n,m)$ in the disk patch.
However, since the structures are stretched along the $y$ direction by
the Keplerian shear, a 1-D FT along $x$ is easier to interpret and
compute.  The power spectrum was thus computed as the average of the
FT of the $M$ rows:
\begin{equation}
  \mathcal{P}(k) = \frac{1}{M}\sum^M_{i=1} \|\mathcal{F}(\Sigma(n,i))\|,
\end{equation}
where $\mathcal{F}$ represents the Fourier transform.  In this case we
chose $N=128$, $M=40$ and we applied an FFT algorithm.
\begin{figure}[t]
  \centering
  \includegraphics[width=8.5cm]{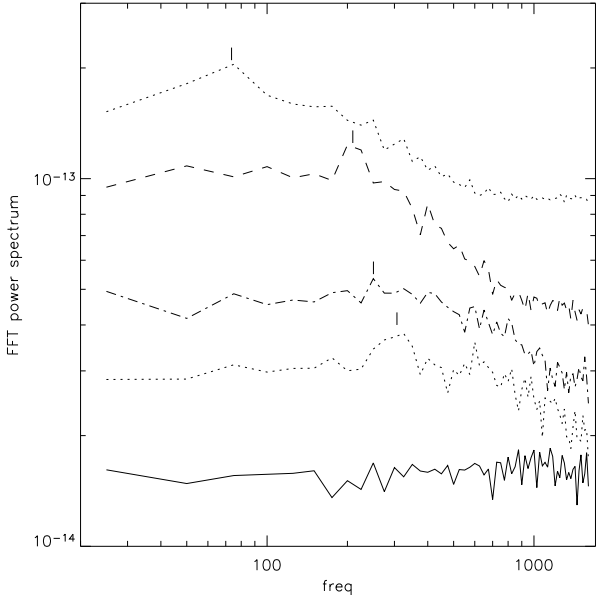}
  \caption{Average power spectrum along $x$.  Each curve
    corresponds to a different time of the simulation, i.e., from
    bottom to top: $t=0$, $t=0.11$, $t=0.15$, $t=0.30$, $t=0.37$ orbit
    (to be compared with the first frames of Fig.~\ref{F:nogas_cm1}).
    The small vertical bar indicates the approximate position of the
    maximum of each curve.  The frequency is expressed as
    (1/$\lambda$~AU$^{-1}$).  To simplify comparisons, an arbitrary
    vertical shift has been added to each curve.
    \label{F:fftresume}
  }
\end{figure}
\begin{figure}[h]
  \centering
  \includegraphics[width=8.5cm] {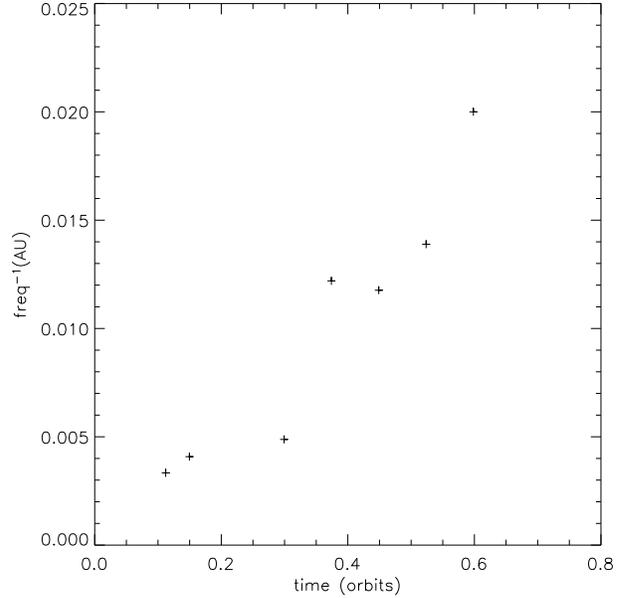}
  \caption{
    Wavelength corresponding to the maximum of the power spectrum, as
    a function of time.
    \label{F:ffttime}
  }
\end{figure}

The results thus obtained are summarized in Fig.~\ref{F:fftresume}.
As expected the power spectrum of the initial conditions corresponds
to white noise.  However, the evolution shows quite well the energy
accumulating in larger and larger wavelengths, as shown by the
displacement of the maximum of the power spectrum.  The last (upper)
curve shows that when large structures are present, a flat spectrum is
recovered at high frequencies, indicating the disappearance of small
scale structures.

The time evolution of the wavelength corresponding to the maximum of
the power spectrum is shown in Fig.~\ref{F:ffttime}.  The noise
associated to the limited resolution does not allow for precise
conclusions to be drawn on the trend.  Qualitatively, all we can
observe is that it is not possible to detect significant departures
from linear growth.

\begin{figure}[h]
  \centering
  \includegraphics[width=8.5cm]{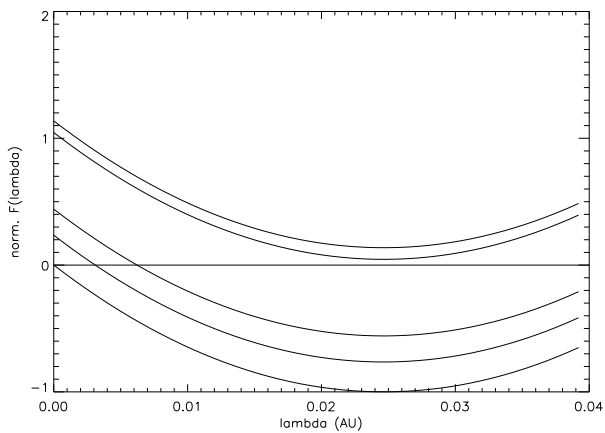}
  \caption{
    The dispersion relation $F(\lambda)$ for the simulation presented
    in Fig.~\ref{F:nogas_cm1}, at the corresponding times.  The lower
    curve corresponds to the initial conditions.
    \label{F:veldisp_flambda_cm1}
  }
\end{figure}

As shown also in Fig.~\ref{F:veldisp_flambda_cm1}, the system evolves
toward an equilibrium state at which velocity dispersions have
stabilized all the wavelengths.  Here, the velocity dispersion $c$
used to compute the dispersion relation (eq.~\ref{dispersion_base}) is
the quadratic average of the three components $c_{x,y,z}$.  Each
component is obtained by the mass-weighted dispersion, i.e., by
computing:
\begin{equation}
  c = \sqrt\frac{\sum_i{m_i (v_i-<\!v\!>)^2}}{\sum{m_i}},
\end{equation}
in which $i$ is the index referred to particles, and sums are computed
over the total particle number.  The initial ordered motion in the
dynamically ``cold'' disk has been transformed in a few orbits into a
warmer swarm of particles, as a result of gravitational stirring.  At
the end of the run, it is clear that the instability is no longer
possible; being favored by the entropy increase, the current state is
stable.

Until now we have not discussed the dissipative effect of inelastic
collisions, resulting in particle growth and mergers.  However, the EC
simulation does not show any difference in comparison to the analysis
shown here.  Collisions and very close encounters thus seem not
relevant for the evolution in the short timescale corresponding to the
lifetime of the observed density waves.  This can also be seen by
considering the timescale due to damping by mutual collisions.
Following Ward (1976), this timescale can be written as:
\begin{equation}
  \tau_{damp} \sim \frac{4}{3}\frac{\rho_{p} R}{\Sigma}\frac{1}{\beta \Omega}
\end{equation}
in which $\beta$ represents the fraction of energy dissipated at each
collision.  In our case, having $\beta=1$, $\tau_{damp}\sim
3\times10^3$ orbits, several orders of magnitudes higher than the
timescale of the velocity dispersion increase observed here.

In other words, the development of the features observed on short
timescales, corresponding to the transition from a dynamically cold,
unstable disk, to a dynamically hot one, is practically independent
of mutual collisions.

\section{The instability in the presence of gas drag}

An efficient dissipation of kinetic energy can be obtained by
introducing friction of the planetesimals from a gaseous nebula.  In
the following, we suppose that the gas flow, represented by the
velocity {\bf u}=($u_x,u_y,u_z$), follows a purely laminar, Keplerian
profile.  In the local co-rotating system this can be written as:
\begin{equation}
  u_x = 0; u_y = -\frac{3}{2}\Omega x; u_z = 0.
  \label{E:keprof}
\end{equation}
The gas drag over the particle is as an additional acceleration having
the form:
\begin{equation}
  \frac{d{\bf v}}{dt} = -\frac{1}{\tau_{stop}}\left({\bf v}-{\bf
      u}(x,y,z)\right),
\end{equation}
where ${\bf u}$ is computed at the position of the particle $(x,y,z)$.
For simplicity of computation, we assume that the stopping time is
proportional to the size of the particle: $\tau_{stop} = R/\gamma$,
$\gamma$ being a constant controlling the amount of gas friction.  A
given value of $\gamma$ will thus correspond to a specific stopping
time for each particle size.

In the discussion following, however, we prefer to express gas drag
through the Stokes parameter, defined as the ratio of the stopping
time to the typical dynamical time of the fluid motion (the orbital
period): $St=\tau_{stop}\Omega/2\pi$.  In our case, we will indicate
the value of $St$ for the initial particle size. 
\begin{figure*}[ht]
  \centering
  \includegraphics[width=13cm]{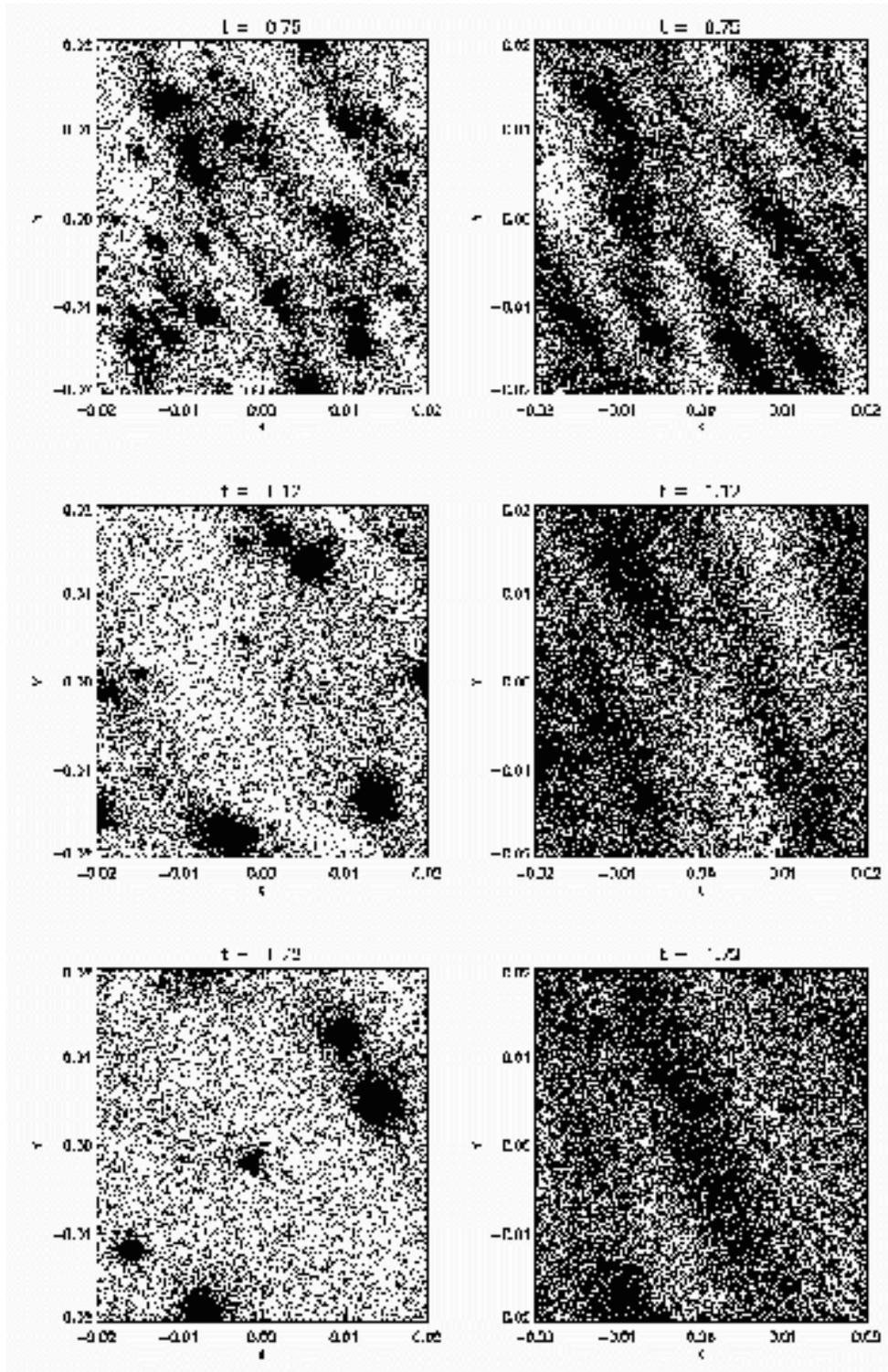}
  \caption{Results from the CM simulation with $St=1$
    (left column) and $St=10$ (right column), with merging particles.
    Large cluster merging is observed in the first case, while in the
    second one only a small spherical cluster survives at the end of
    the simulation.  Circles in the left column snapshots represent
    the area used for computing detailed statistics, as detailed in
    the text.
    \label{F:gamma10-11}
  } 
\end{figure*}

Depending on the value of $\tau_{stop}$, the particles will be more or
less forced to follow the Keplerian shear (\ref{E:keprof}).  On the
other hand, particles exactly moving as in eq.~(\ref{E:keprof}) will
not be influenced by any energy dissipation.  As a result, no radial
migration can be reproduced in this ideal case, and the only effect of
the drag is to reduce velocity dispersions.

In the case of the minimum mass solar nebula, the gas drag would
hardly have any effect on planetesimals with $R=10$~km in the time
span covered by the simulation presented here.  Instead of imposing a
realistic drag, we thus look for the smallest dissipation necessary to
preserve some structures in the disk.  Since the lifetime of the
structures observed in the previous section is about $\sim$1 orbit, we
predict that a stopping time of the same order of magnitude should
deeply affect their evolution.  We thus test the evolution assuming
$St=1$.

In this case the effects of gas drag are readily apparent.  Both in CM
and EC simulations, the initial amplification of small wavelength
fluctuations is followed by their hierarchical growth, as usual.  The
dissipation, however, prevents high-density regions from dispersing,
and the mutual interactions make clusters grow by merging.  As can be
seen qualitatively from the time evolution of the $(x,y)$ distribution
of particles (Fig.~\ref{F:gamma10-11}, left column), the clusters
evolve toward circular symmetry.

Except for the very beginning of the dynamical evolution, this kind of
distribution is no longer well suited for characterization by the
study of the Fourier transform, since a periodic density wave is not
present.  On the other hand, the two-point correlation function can be
computed in order to give a statistical estimate of the degree of
clustering.  Formally, the two-point correlation function $\xi(r)$ is
defined as the excess probability (relatively to a homogeneous
distribution) of finding a point in a volume $dr^s$ ($s$ being the
space topological dimension) at distance $r$ from a randomly chosen
point . If $\delta P$ is the probability of finding a particle in the
volume $dr^s$, then:
\begin{equation}
  \delta P = \bar{n} \left[ 1 + \xi(r) \right] dr^s,
\end{equation}
in which $\bar{n}$ is the average number density of points in the
distribution.  Since we are mainly interested in the distribution on
the $z=0$ plane, the computation of $\xi(r)$ was made in two
dimensions ($s=2$), by considering the positions of the particles
projected on that plane.

\begin{figure}[ht]
  \centering
  \includegraphics[width=8.5cm]{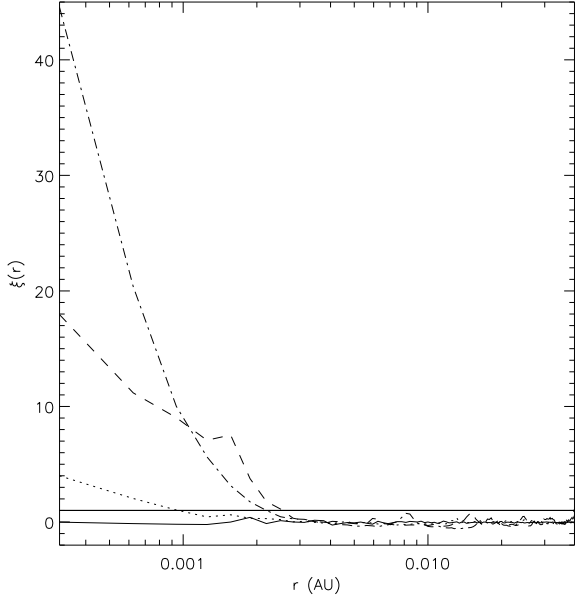}
  \caption{The two-point correlation function at times $t=0$,
    $t=4.70$, $t=7.28$, $t=10.81$ orbits.  The line $\xi(r)=1$ has
    been added as a reference.
    \label{F:corr2p}
  } 
\end{figure}

Conventionally the length $\lambda_h$ for which $\xi(\lambda_h) \equiv
1$ is assumed to be the ``characteristic length,'', i.e., the scale of
the largest typical clusters.  From a practical point of view, we
computed $\xi(r)$ by the method discussed in~\cite{Rivolo}.  Since
boundary conditions are periodic, the procedure is exact in our case.
The results obtained for four epochs are illustrated
in~Fig.\ref{F:corr2p}.  The strong growth of $\xi(r)$ at small
distances is a clear sign of the increasing clustering.  It can be
noted that the probability of finding a particle close to another one
(at a vanishing distance) grows to more than 40 times the average
value.  This value is averaged over the whole patch, thus including
those particles that are not in the clusters.  Nevertheless, since the
fraction of clustered bodies is dominating the distribution, this can
be considered to be a reliable estimate, at least at the end of the
simulation.  In the time interval considered, the characteristic
length stabilizes around $\lambda_h \sim 2.5\times10^{-3}$~AU.

Taking this value for the average radius of the clusters, we can
estimate their overall properties.  At the latest time shown in
(Fig.~\ref{F:gamma10-11}, left column), the cylinders of radius
$\lambda_h \sim 2.5\times10^{-3}$~AU centered on the two larger
clusters (in the upper right quadrant of the patch) include masses of
$2.3\times10^{22}$~g and $3.6\times10^{22}$~g.  Both clusters include
a member with a diameter larger than 100 km in their cores.  If the
entire mass was concentrated into single bodies of volume density
1~g/cm$^3$, they would have radii of 176 and 204~km, 
respectively.  
The corresponding Hill radii would be 4.7$\times10^{-3}$~AU and
5.4$\times10^{-3}$~AU.  At later times in the simulation the two
clusters merge into a single one, containing about 40\% of the total
mass available in the patch.  At the time reached by our longest
simulations (i.e., t$\sim$4~orbits) only two large clusters remain,
whose average distance is larger than their Hill radius.  However, we
do not present here a detailed analysis of such end states, since to
study the dynamics of such large clusters a much larger patch (of
several Hill radii per side) would be necessary.  We stress again here
that the same qualitative behavior, on the same timescale, is observed
when particle collisions are treated as bounces (EC case) or as
merging events (CM).
\begin{figure}[h]
  \centering
  \includegraphics[width=8.5cm]{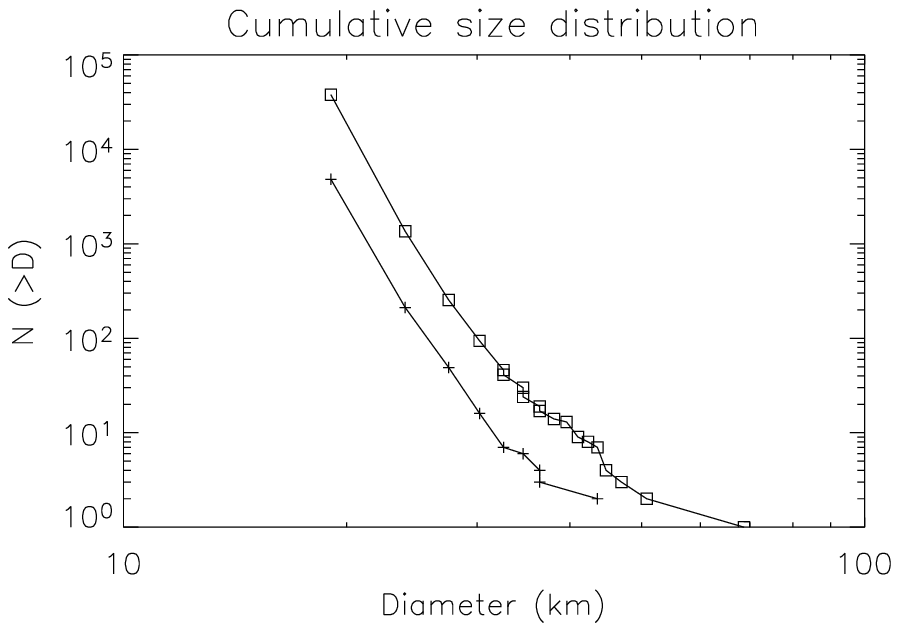}
  \includegraphics[width=8.5cm]{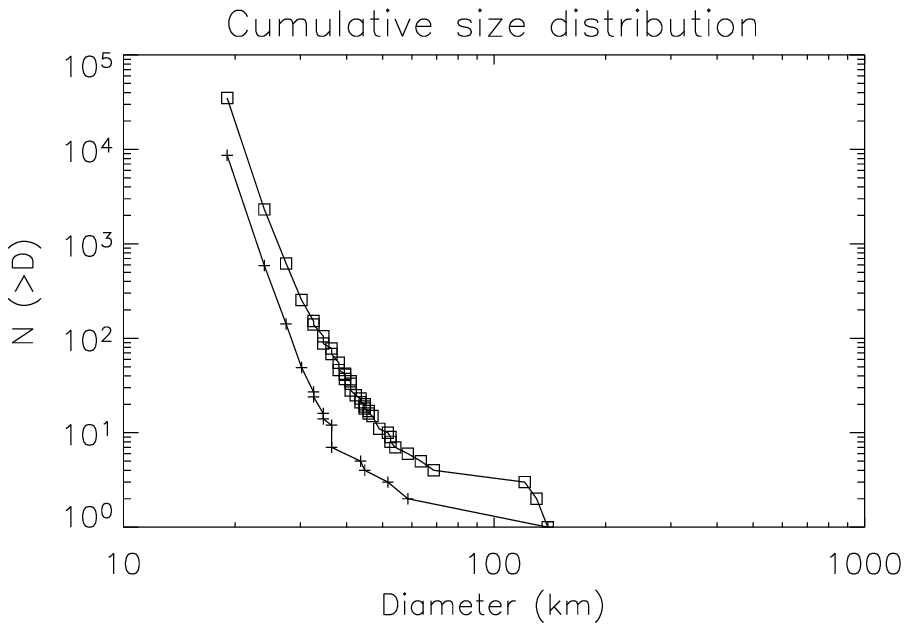}
  \includegraphics[width=8.5cm]{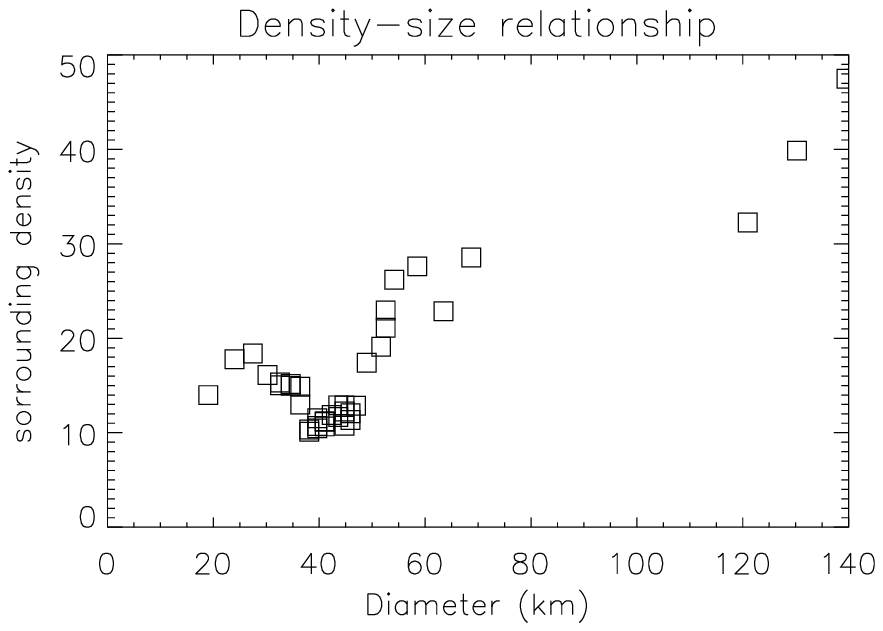}
  \caption{Some statistics concerning the size distribution of
    the simulation in Fig.~\ref{F:gamma10-11} (left column).  The
    upper and middle panels show the size distribution at $t=7.05$ and
    $t=7.15$ orbits, respectively.  The upper curve is the
    distribution obtained when including all particles.  The lower
    curve is the distribution referred to the domain inside the
    corresponding circle in Fig.~\ref{F:gamma10-11} (lower left
    panel).  The plot of particle size {\it vs}.\ local density (lower
    panel), shows the tendency of particles to grow faster in
    overdense regions, and is referred to time $t=10.81$ orbits.
    \label{F:gamma10-10size}
  }
\end{figure}

In this last case it is interesting to check the properties of the
evolving cumulative size distribution of particles.
Fig.~\ref{F:gamma10-10size} presents this distribution at two
different times.  Besides the global distribution, the size
distribution restricted to a specific cluster is also computed.  It
can be seen that (especially at the most evolved stages) large members
are present inside the clusters.  We also computed the volume density
on a cubic grid $N\times M\times L$ with $N=M=L=32$ and used it to
study the relation between the size of a particle and the density of
the cell in which it is contained.  The existing correlation
(Fig.~\ref{F:gamma10-10size}, lower panel) indicates that large
particles form and remain inside dense areas.  We can thus state that
particle growth is acting preferentially inside clusters.

Concerning some specific cluster properties, we note that: 
\begin{itemize}
\item[a--] The cluster shape is close to spherical.  A plot of
  isodensity surfaces shows that a small flattening is indeed present
  at a distance of $\sim$2$\times10^{-3}$ from the cluster center, but
  is negligible in its inner part.
\item[b-] The density profile in the central part of the cluster
  behaves like $\rho\sim r^{-2}$ (Fig.~\ref{F:radprof}).  This central
  region is about 10$^{-3}$AU in size.
  \begin{figure}[h]
    \centering
    \includegraphics[width=8.5cm]{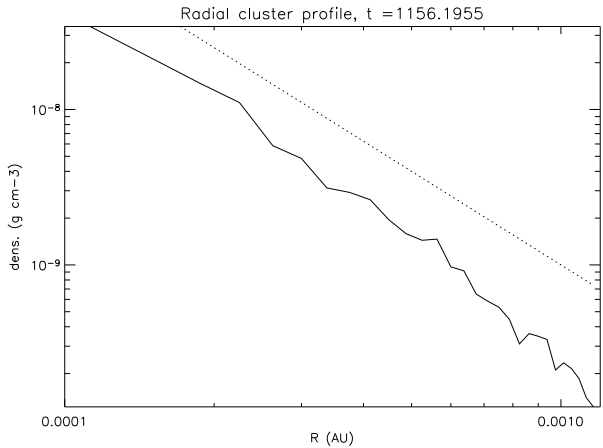}
    \caption{The average radial density profile (solid line) in the most dense
      part of the cluster circled in Fig.~\ref{F:gamma10-11} (middle
      panel, left column).  The dotted line represents the power law
      $R^{-2}$.
      \label{F:radprof}
    }
  \end{figure}
\item[c--] The orientation of the orbits around the cluster barycentre
  appear to be randomized.  This can be seen by the fact that the
  velocities do not show any systematic orientation.  We have analyzed
  the horizontal velocity {\bf v$_h$} after having subtracted the
  average Keplerian shear of each particle and having decomposed it
  into its radial and tangential components (v$_r$, v$_t$) relative to
  the direction of the cluster barycentre.  No trend can be identified
  by plotting v$_r$ or v$_t$ as a function of $r_{h}$ (distance from
  barycentre as projected on the disk equatorial plane).  In addition,
  the direction of v$_h$ is uniformly distributed in the [0,2$\pi$]
  range.  These statistics do not change if they are computed by
  considering all particles or only a subset of them, close to the
  disk average plane.  The typical velocity dispersions close to the
  cluster center are very small, of the order of
  $\sigma_{v_r}\sim\sigma_{v_t}\sim 1.5$~m~s$^{-1}$.
  \begin{figure}[h]
    \centering
    \includegraphics[width=8.5cm]{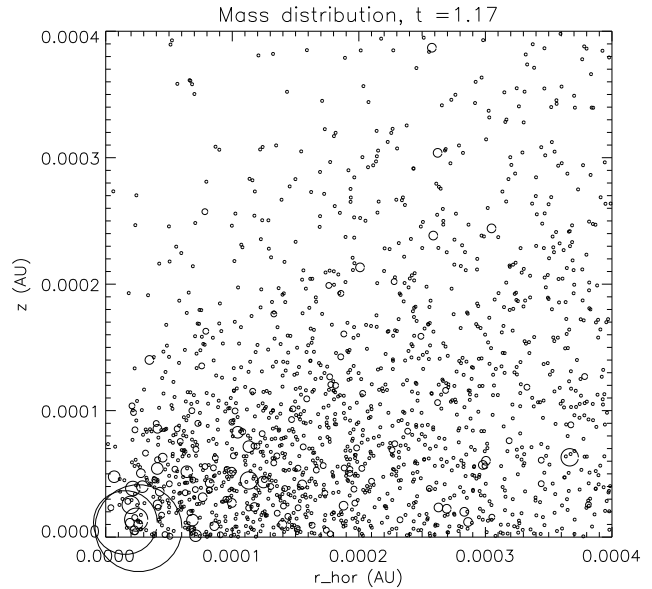}
    \caption{The particle distribution in the most dense part of the cluster 
      circled in Fig.~\ref{F:gamma10-11} (middle panel, left column).
      Each circle is centered on the position of a particle in the
      $(r,z)$ plane; the radius is {\it proportional to the mass of
        the particle} in order to emphasize the effect of size
      differences.  As can be seen, the center of the cluster is
      populated by large bodies.
      \label{F:xzmass}
    }
  \end{figure}
\item[d--] In the CM case, particle sizes are roughly sorted
  accordingly to their distance to the cluster center: large particles
  are close to the barycentre, while small bodies are scattered all
  around (Fig.~\ref{F:xzmass}).  This behavior can be expected as the
  result of a sort of energy equipartition due to dynamical friction.
  We recall that, in flattened, stable disks, the effect of dynamical
  friction is to reduce the velocity dispersion of large bodies that
  are surrounded by a swarm of small particles.  As a consequence,
  this swarm is ``heated'' and its velocity dispersion increases.
  From the point of view of the vertical distribution relative to the
  average plane, this translates into the large particles being less
  dispersed than the small ones.  Inside the clusters studied here,
  the mechanism is the same, except that the system has a spherical
  symmetry.
  
\item[e--] According to these dynamics, runaway growth is present,
  with the largest body detaching from the distribution and growing
  faster.
\end{itemize}

It must be noted that both the radial profile and the randomization of
velocities are in agreement with that of a self-gravitating,
completely virialized cluster.  The $\rho\sim r^{-2}$ density
distribution was suggested for virialized halos with isothermal
profiles (\cite{gunngott}).  Indirectly, the virialization is
confirmed by the fact that, given the order of magnitude of the
cluster size, the typical cluster crossing time for a particle is
$t_{cross}\sim2.5\times10^{-3}$~orbits, i.e., several times less than
the cluster lifetime observed in our simulations.  The corresponding
two-body relaxation time for globular clusters composed of $N_{clus}$
bodies is usually computed as $t_{rel}\sim0.1\frac{N_{clus}}{ln\ 
  N_{clus}}t_{cross}$.  In our most populated clusters,
$N_{clus}\sim10^4$, yielding $t_{rel}\sim0.64$~orbits, i.e., less than
the duration of our simulations.

Furthermore, it must be noted that the clusters interact and merge on
a timescale of a few orbits.  During this interval, the dynamics is
not affected by collisions, being comparable both in the CM and the EC
scenario.

We also verified that $St\sim1$ corresponds to the gas drag that
maximally preserves the structures, by running some simulations with
higher and lower values of $\gamma$, in power-of-ten steps.  In
general, we observe that with a very strong drag ($St=10^{-3}$), as
expected, all velocity dispersions are rapidly damped.  The vertical
distribution becomes very thin and close encounters are favored by the
nearly two-dimensional distribution.  Due to small relative velocities
and increasingly strong gravitational focusing, mutual collisions are
frequent. If merging is allowed, the masses of the particles grow
fast.  This particular simulation ends at $t=3800$ with $N$ decreased
to $N=980$, and without the formation of any structures.

\begin{figure}[h]
  \centering
  \includegraphics[width=8.5cm]{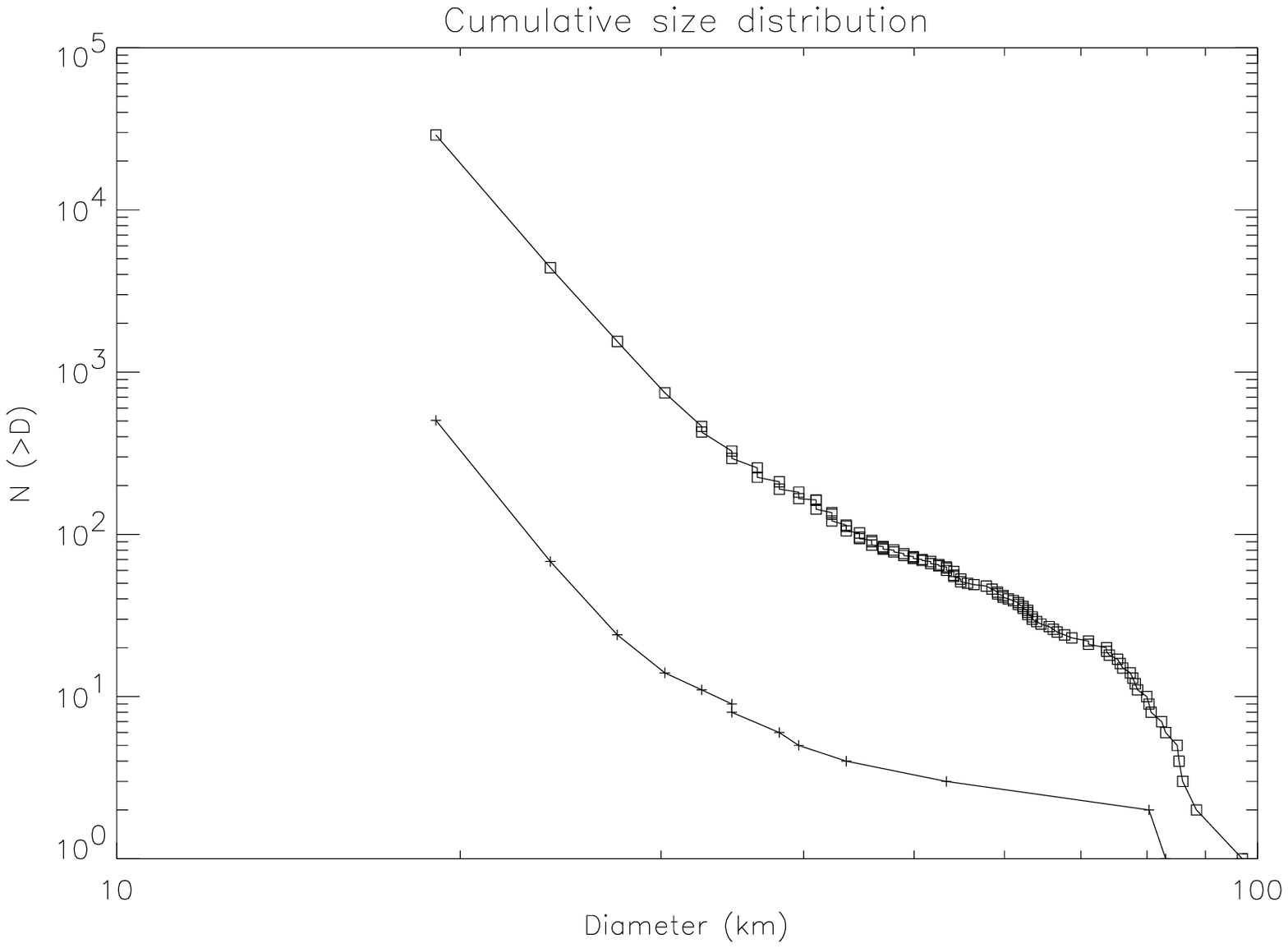}
  \includegraphics[width=8.5cm]{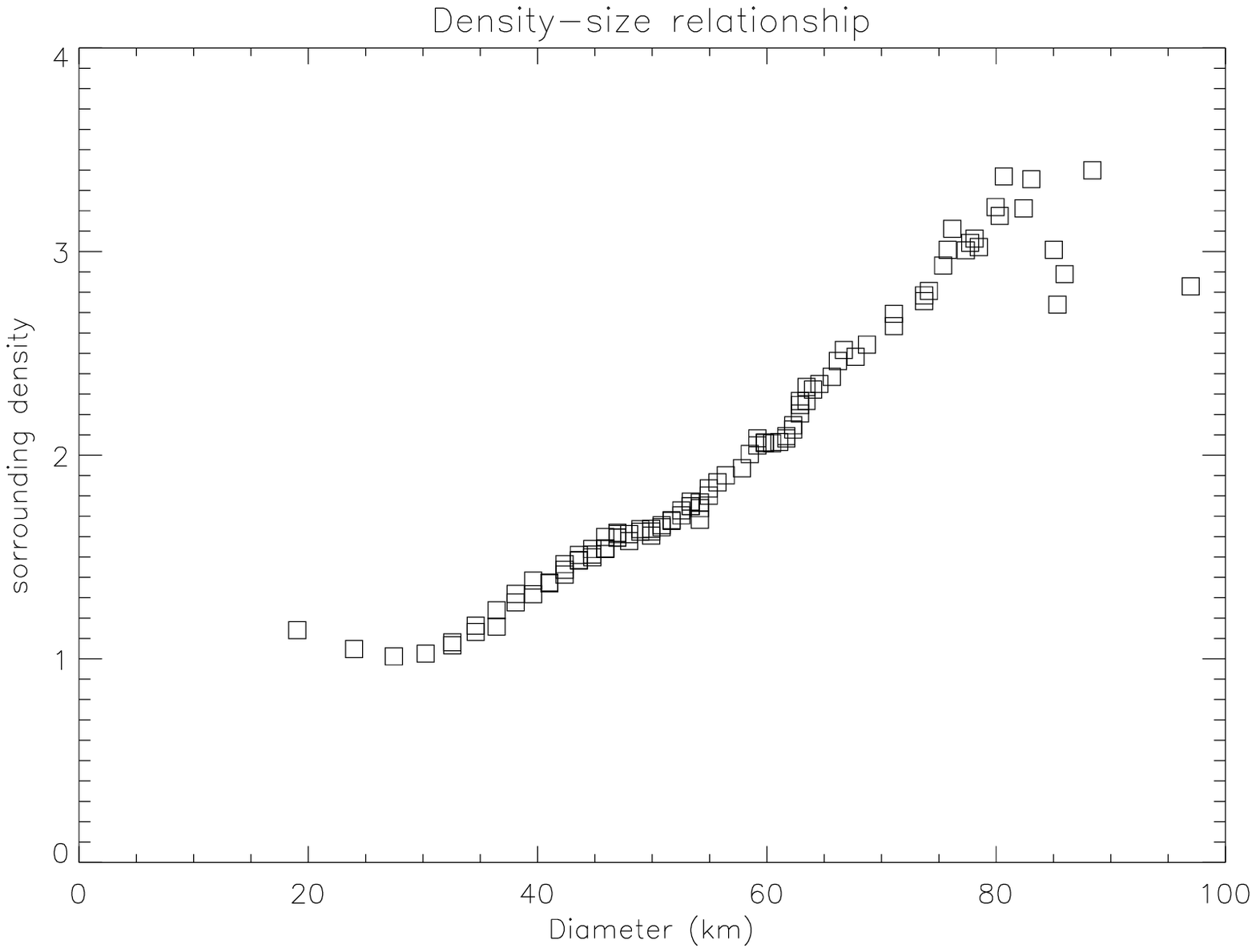}
  \caption{Some results from the CM simulation with
    $St=0.1$.  The upper panel shows the final cumulative size
    distribution: the upper curve represents the distribution computed
    by taking into account all the particles.  The lower curve takes
    into account only particles belonging to the larger cluster.  The
    bottom panel shows that a correlation between the size of the
    particle and the surrounding volume density is present.
    \label{F:gamma109dist}
  }
\end{figure}

Only at $St = 0.1$ do clusters of particles begin to emerge from the
distribution.  The velocity dispersion increases with time, and
accretion is still very efficient.  However, a large fraction of
particles remains outside the clusters.  It can be seen that a strong
correlation exists between the size of the particles and the
surrounding volume density of solids (Fig.~\ref{F:gamma109dist}).

On the other hand, for larger stopping times ($St=10$) it is
interesting to note that the transient formation of several clusters
is present.  However, instead of merging, they rapidly disappear due
to gravitational stirring.  In the case of our simulation, we observe
one remaining condensation (containing only about 2.3\% of the total
mass available), indicating that cluster survival is at least
marginally allowed (Fig.~\ref{F:gamma10-11}, right column).

Even longer stopping times will yield a drag insufficient to promote
cluster formation.

We also tested the formation of clusters starting with completely
different initial conditions, i.e. with the ``hot disk''
shown in Fig.~\ref{F:nogas_cm1} (bottom right panel), and allowing the gas to operate the
``cooling'' necessary to drive the system to the unstable regime. Some
differences in the details of the evolution exist. In particular, due to
the higher velocity dispersion, the instability does not take place at
small scales, but directly at a scale close to $\lambda_c$. Details of
this process will be further investigated; here we just want to stress
that, even in this ``inverse'' case, clusters form and have the same 
properties as presented in this work.

\section{Conclusions}

The system explored, when no dissipation is acting, evolves
spontaneously toward a stable configuration in which velocity
dispersions are too high to allow any gravitational instability.  This
is clearly shown in the simulation presented in
Fig.~\ref{F:nogas_cm1}.  As already discussed, mutual collisions are
not able to damp velocity dispersions efficiently, and play a
secondary role over the observed dynamical evolution.  However, this
can be regarded as a consequence of the particular regime that has
been chosen.  In fact, closer to the sun and/or using much smaller
particles, the dynamics could be collision-dominated (\cite{furuya}).

Our simulations underline the role of both the non-linear evolution of
the collapsing layer and of its granularity.  Thanks to the
hierarchical growth of structures, the particle-particle interaction
is gradually substituted by the cluster-cluster one at larger scales.
At the end of the simulations that include gas drag, large clusters
behave as super-particles dominating the dynamics.  A first set of
tests, made by isolating a single cluster from the surrounding
particles and computing its evolution on long times, seems to show
that the clusters are intrinsically ``stable'' structures.  Gas
friction also guarantees that they can shrink and collapse to form
large bodies, although on time scales much larger than those explored
here.

Of course, our approach is also affected by some limitations.  A
strong limit of our study concerns the use of a layer of equally sized
particles representing the solid component.  It has been suggested
that a full range of particle sizes, exhibiting different couplings to
the nebular gas, could deeply affect the instability of the disk
(\cite{Weiden95}, \cite{ward00}).  This point will be investigated in
a coming paper.

Furthermore, in order to study the dynamics of larger bodies in the
presence of dissipation, we introduced a gas drag law with intensity
not physically related to particle sizes.  In other words, the
resulting stopping time was much smaller than any realistic value that
could correspond to km-sized planetesimals in the environment of a
classical protoplanetary disk.

On the other hand, it must be noted that the formation of
self-gravitating, virialized clusters is not specific to this choice.
In fact, the dynamics observed is essentially collisionless, being
unaltered by the presence of merging or perfectly bouncing particles.
Thus, as commonly done in cosmology, we can assume that our discrete
distribution of large bodies just represents a single possible
sampling of the density field.  Probably, its dynamics would be
qualitatively equivalent if the field was sampled by much smaller
bodies, as long as long-range interactions are possible (i.e., the
Hill radius is much larger than the physical radius) and the stopping
time is comparable to the orbital period (St$\sim$1).  In a
``realistic'' minimum-mass nebula, this could be true for particle
sizes of the order of $R\sim 1$~m, sufficiently far from the sun (in
order to satisfy the Hill radius constraint), provided the radial
velocity dispersion can be neglected---which would not always be the
case.  Under these conditions, the mechanism illustrated here could
thus dominate the dynamics of an unstable layer of solids.  Further
investigations are under way following this direction, and will be the
subject of a future publication.

Another major drawback in the direct application of this scenario to
the growth of planetesimals is related to the fact that the action of
particles over gas is neglected.  In fact, it is clear that, when
solids are strongly concentrated inside clusters, they could deeply
affect the motion of the surrounding gas.  This feedback was already
invoked as a cause of stabilization of the midplane dust layer,
through the stirring operated by the turbulence.  In that scenario,
the main cause of turbulence is the shear due to the systematic
velocity difference between dust particles and gas.  In our case,
since cluster member velocities are randomized, no net systematic
motion would be induced in the average gas flow.  However some kind of
small-scale turbulence, with the associated diffusivity, could oppose
the collapse.
 
In conclusion, we want to underline the observation, in our numerical
simulations, of the extremely rich dynamics hidden behind
gravitational instability in a Keplerian disk.  Even though the
application to the study of planetesimal growth requires further
effort, the phenomena exposed here may always be present when
collective effects, rather than collisional effects, dominate the
dynamics.

\begin{acknowledgements}
  This study was supported in part by the {\it Programme Nationale de
    Plan\'etologie}.  D. Richardson acknowledges support of the NASA
  Origins of Solar Systems contract \#NAG-511722.  S.J.
  Weidenschilling was supported by the NASA Planetary Geology and
  Geophysics Program, Grant \#NAG5-13156.  We thank the IDRIS
  computing center facility for the use of the SP3 and SP4 parallel
  computers.  SIVAM and SIVAM II computing systems at the OCA were
  also used for this work.  We wish also to acknowledge useful
  discussions with P.-H. Chavanis and G. Murante, and the work of an
  anonymous referee.
\end{acknowledgements}


\begin{thebibliography}{}
  
\bibitem[Barge et al. 1995]{barge95} Barge, P., \& Sommeria, J. 1995,
  A\&A, 295(1), L1
  
\bibitem[Edgeworth 1949]{edgeworth49} Edgeworth, K.E. 1949, MNRAS, 109, 600
  
\bibitem[Furuya et al. 2002]{furuya} Furuya, I., Daisaka, H.,
  Nakagawa, Y. 2002 in The Proceedings of the IAU 8th Asian-Pacific
  Regional Meeting (Tokyo), 2, 21
  
\bibitem[Goldreich \& Ward 1973]{Golward} Goldreich, P., \& Ward, W.R.
  1973, ApJ, 183, 1051

\bibitem[Goodman et al. 1980]{Goodman} Goodman, J., \& Pindor, B. 
2000 Icarus, 148, 537

\bibitem[Gunn \& Gott 1972]{gunngott} Gunn, J.E., Gott, J.R. 1972,
  ApJ, 176, 1
  
\bibitem[Richardson 1994]{derek94} Richardson, D.C. 1994, MNRAS, 269(2),
  493
  
\bibitem[Richardson et al. 2000]{Derek} Richardson, D.C., Quinn, T.,
  Stadel, J., \& Lake, G. 2000, Icarus, 143, 45
  
\bibitem[Rivolo, 1986]{Rivolo} Rivolo, A.R. 1986, Astron. J., 301, 70
  
\bibitem[Safronov 1969]{Safronov} Safronov, V.S. 1969, Evolution of
  the protoplanetary cloud and formation of the Earth and the planets
  (Nauka, Moskow).
  
\bibitem[Tanga et al. 1996]{tanga96} Tanga, P., Babiano, A., Dubrulle,
  B., \& Provenzale, A. 1996, Icarus, 121, 158
  
\bibitem[Tanga et al. 2002]{tanga02} Tanga, P., Michel, P.,
  Richardson, D.C. 2002, A\&A, 315, 613
  
\bibitem[Ward 1976]{ward} Ward, W.R. 1976, in Frontiers of
  Astrophysics, ed E.H.Avrett, (Harvard Univ. Press, Cambridge)
  
\bibitem[Ward 2000]{ward00} Ward, W.R. 2000, in Origin of the Earth
  and Moon, ed.  R.M Canup and K. Righter (Univ. of Arizona Press,
  Tucson)
  
\bibitem[Wisdom \& Tremaine 1988]{wt88} Wisdom J., Tremaine, S., 1988,
  Astroph. J., 95, 925
  
\bibitem[Weidenschilling 1980]{Weiden80}  Weidenschilling, S. J. 1980, 
  Icarus 44, 172

\bibitem[Weidenschilling 1995]{Weiden95} Weidenschilling, S.J. 1995,
  Icarus 116, 433
  
\bibitem[Weidenschilling 1997]{Weiden97} Weidenschilling, S.J. 1997,
  Icarus 127, 290
  
\bibitem[Weidenschilling 2003]{Weiden03} Weidenschilling, S.J. 2003,
  Icarus 165, 438
  
\bibitem[Youdin 2002]{Youdin02} Youdin, A.N., Shu, F.H., ApJ, 580, 494

\end{thebibliography}
\end{document}